\documentclass[twocolumn,aps,prc,10pt,superscriptaddress,preprintnumbers]{revtex4-1}

\usepackage[english]{babel}
\usepackage[dvipsnames,svgnames,x11names]{xcolor}
\usepackage[markup=underlined]{changes}
\usepackage{float}
\restylefloat{table}

\usepackage{graphicx}
\usepackage{amssymb}
\usepackage{amsmath}
\usepackage{epsfig}
\usepackage{bm}
\usepackage[dvipsnames]{xcolor}
\usepackage{multirow}
\definechangesauthor[color=blue]{GP}
\definechangesauthor[color=magenta]{MN}
\definechangesauthor[color=red]{MB}
\definechangesauthor[color=green]{MK}
\definechangesauthor[color=purple]{*}

\colorlet{Changes@Color}{black!90}

\begin{document}

\preprint{J-PARC-TH-0256}

\title{Critical net-baryon fluctuations in an expanding system}

\author{Gregoire Pihan}
\email{gregoire.pihan@subatech.in2p3.fr}
\affiliation{SUBATECH UMR 6457 (IMT Atlantique, Universit\'e de Nantes, IN2P3/CNRS), 4 rue Alfred Kastler, 44307 Nantes, France}
\author{Marcus Bluhm}
\affiliation{SUBATECH UMR 6457 (IMT Atlantique, Universit\'e de Nantes, IN2P3/CNRS), 4 rue Alfred Kastler, 44307 Nantes, France}
\author{Masakiyo Kitazawa}
\affiliation{Department of Physics, Osaka University, Toyonaka, Osaka 560-0043, Japan}
\affiliation{J-PARC Branch, KEK Theory Center, Institute of Particle and Nuclear Studies, KEK, 203-1, Shirakata, Tokai, Ibaraki, 319-1106, Japan}
\author{Taklit Sami}
\affiliation{SUBATECH UMR 6457 (IMT Atlantique, Universit\'e de Nantes, IN2P3/CNRS), 4 rue Alfred Kastler, 44307 Nantes, France}
\author{Marlene Nahrgang}
\affiliation{SUBATECH UMR 6457 (IMT Atlantique, Universit\'e de Nantes, IN2P3/CNRS), 4 rue Alfred Kastler, 44307 Nantes, France}

\begin{abstract}
	In this work we study the consequences of a longitudinal Bjorken expansion and a Hubble-like temperature cooling scenario on a 1+1D non-linear model of the diffusive dynamics of fluctuations in the net-baryon density. The equilibrium behavior of the fluctuations is fully encoded in the temperature dependence of the susceptibilities on the crossover side both in the vicinity of the assumed location of the critical point and at vanishing baryo-chemical potential in-line with lattice QCD calculations. We demonstrate the great sensitivity of the fluctuation observables on the dynamics, in particular on the diffusion length and the freeze-out conditions. While the critical signals are visible and the critical region is broadened by the expansion, a too small diffusion length can strongly reduce the amplitude of the signals. We propose to search for significant anti-correlations of baryons at intermediate rapidity experimentally and to map out the rapidity dependence of the fourth-order cumulant, which in the presence of a critical point (and only in its presence) has a pronounced minimum at intermediate rapidities.
\end{abstract}

\maketitle

\section{Introduction}
\label{sec:Intro}

Near a critical point the conventional relativistic fluid dynamical description ceases to be valid as thermal fluctuations, which are usually assumed to be negligible, have an impact at the macroscopic scale of a system. The dynamical behavior of these critical fluctuations has gained recent attention in the context of mapping the phase diagram of quantum chromodynamics (QCD) by means of heavy-ion collisions and looking for signals of a potential critical point~\cite{Nahrgang:2016ayr,Luo:2015doi,Bzdak:2019pkr,Bluhm:2020mpc,An:2021wof}. 

For the study of the fluctuation dynamics the tool of fluid dynamics is particularly well suited, because its deterministic version describes well the event-averaged bulk observables in heavy-ion collisions~\cite{Hirano:2008hy,Romatschke:2017ejr} and the critical mode is a fluid dynamical field~\cite{Hohenberg:1977ym}. Currently there are two different approaches which are followed to include fluctuations into fluid dynamics: the first is a hydro-kinetic approach, where the correlator of the critical mode is solved in addition to the fluid dynamical equations~\cite{Akamatsu:2016llw,Stephanov:2017ghc,An:2019osr,Rajagopal:2019xwg,An:2019csj,Du:2020bxp}. The resulting set of coupled equations is then deterministic. The second approach, which we follow up on in this work, is stochastic fluid dynamics, where an explicit stochastic noise term is included in the fluid dynamical equations which are then solved event-by-event~\cite{Nahrgang:2011mg,Herold:2016uvv,Murase:2016rhl,Nahrgang:2017oqp,Sakaida:2017rtj,Bluhm:2018plm,Singh:2018dpk,Nahrgang:2018afz,Bluhm:2018qkf,Nahrgang:2020yxm}. For a comparison of the two approaches see~\cite{Bluhm:2020mpc,De:2022vlm}. The evolution of fluctuations in simple diffusive systems is also studied in~\cite{Kitazawa:2013bta,Kitazawa:2015ira}.

Heavy-ion collision experiments at the LHC, the RHIC, at GSI or the future FAIR, are expected to measure critical fluctuations in the event-by-event fluctuations of particle multiplicities related to conserved charges of QCD~\cite{Asakawa:2015ybt}, such as net-proton number fluctuations~\cite{STAR:2020tga,HADES:2020wpc}. By varying the beam energy $\sqrt{s_{N N}}$ regions with different baryo-chemical potential can be probed and coming close to the potential critical point the fluctuation observables should grow large. So far predictions have been formulated based on thermodynamic calculations~\cite{Stephanov:1998dy,Stephanov:1999zu}.

However, a direct comparison between theoretical predictions at thermal equilibrium and experimental measurements is not possible. In fact, the in- or out-of-equilibrium situation of the fluctuations is not known a priori. Thermal fluctuations must diffuse over the entire system before equilibrium can be reached. In heavy-ion collisions, the impact of the violent longitudinal expansion on the diffusive dynamics of the fluctuations as well as the small lifetime of the medium of the order of $10$~fm/c may prevent fluctuations to diffuse sufficiently and, thus, any equilibrium situation to occur both during and at the end of the evolution. This effect is largely enhanced for fluctuations near the critical point due to the critical slowing down e.g.~the downscaling of the diffusion length in the vicinity of the critical point~\cite{Berdnikov:1999ph}. 

The present study aims at investigating the competition between the diffusion and the expansion in heavy-ion collisions and its impact on the dynamics of the critical fluctuations of the net-baryon density. To achieve this, we use the prescriptions in~\cite{Nahrgang:2018afz} and~\cite{Sakaida:2017rtj}. In~\cite{Nahrgang:2018afz} a stochastic diffusion equation (SDE) for the net-baryon density fluctuations is studied, where the criticality is encoded in a free-energy functional containing non-linear contributions. In~\cite{Sakaida:2017rtj} the stochastic diffusion is solved analytically in the linear limit for a Bjorken expansion. Here, we revisit the construction of the free-energy functional via the temperature parametrization of its second and fourth order susceptibilities. This allows us to smoothly connect the equilibrium properties in the scaling region coming from a mapping of the 3D Ising model into the QCD phase diagram with lattice QCD calculations at vanishing baryo-chemical potential. The parametrization of the susceptibilities also permits to fully incorporate the expansion scenario into the equilibrium properties. 

We first proceed to compare our results to the analytical expectations obtained in the limit of a linearized free-energy functional containing only a Gaussian second-order coupling. This step is important for any stochastic formulation of the dynamics. Then, for the full nonlinear model we study the time evolution, the dependence on the diffusion length and the rapidity window dependence of the second and fourth-order fluctuation observables.

\section{Stochastic diffusion equation for the net-baryon density}
\label{sec:II}

In this section, we derive the stochastic diffusion equation and the corresponding noise correlator for a rapidly expanding medium as created in ultra-relativistic heavy-ion collisions. In this case, Milne coordinates represent a convenient choice. Then, we connect the derived evolution equation with known properties of the phase transition between the quark gluon plasma (QGP) and the hadronic phase. 

\subsection{Stochastic diffusion in Milne coordinates}
\label{sec:IIA}

We start from the conservation equation for the net-baryon number current four-vector $N^\mu$ 
\begin{equation}
    \partial_{;\mu}N^{\mu}=0 \,,
    \label{eq:N}
\end{equation}
where $N^{\mu}$ can be expressed as ${N^{\mu}=n_Bu^{\mu}+j^{\mu}}$~\cite{Hirano:2008hy} and $\partial_{;\mu}$ is a covariant derivative. Here, $n_B$ is the net-baryon density and the conventionally deterministic, dissipative current $j^\mu$ receives an additional stochastic current contribution $\xi^\mu$ in this work following 
\begin{equation}
        j^{\mu} = \sigma T \Delta^{\mu\nu} \partial_{\nu} \left\{ \dfrac{\mu_B}{T} \right\} + \xi^{\mu} \,. 
        \label{eq:current}
\end{equation}
This defines $N^\mu$ in the basis $(u^{\mu},\Delta^{\mu\nu})$, where $u^\mu$ is the fluid four-velocity and $\Delta^{\mu\nu}=g^{\mu\nu}-u^{\mu}u^{\nu}$ is the projection operator orthogonal to $u^\mu$. We use the metric tensor $g^{\mu\nu}=$~diag$(+, -, -, -)$. In Eq.~\eqref{eq:current}, $\mu_B$ is the baryo-chemical potential, $T$ is the temperature and $\sigma$ is a mobility coefficient. We will assume that the components of the stochastic current are Gaussian white noise where, in line with the deterministic part of the dissipative current in Eq.~\eqref{eq:current}, the noise-noise correlators are local in space and time and given as~\cite{Kapusta:2012gm} 
\begin{equation}
        \langle \xi^{\mu}(X) \,\xi^{\nu}(X') \rangle = - 2 \sigma T  \delta^{(4)}(X-X') \Delta^{\mu \nu}\,.
        \label{eq:correlator}
\end{equation}
This guarantees that the fluctuation-dissipation balance is locally respected such that the relative importance of the fluctuations depends on the value of the transport coefficient. 

Inserting Eq.~\eqref{eq:current} into Eq.~\eqref{eq:N} gives rise to a stochastic diffusion equation which we discuss in the following. In addition, we define the net-baryon number diffusion length as $D = \sigma T/n_c$, where $n_c$ is a constant critical net-baryon density which we set, in Cartesian coordinates, to $n_c=1/3$~fm$^{-3}$ throughout this work, and assume that the temperature cools spatially homogeneous during the fireball evolution. 

In our study, we model the dynamics of the medium by a Bjorken-type expansion~\cite{Bjorken:1982qr}, i.e.~we consider boost-invariance in the longitudinal (collision axis) direction of the created fireball. This represents a good approximation for the mid-rapidity region at large collision energies. 
In this model, the fluid four-velocity reads neglecting the transverse dynamics 
\begin{equation}
    u^{\mu}= \dfrac{1}{\tau}(t, 0, 0, z) \,.
    \label{eq:Bjorkenflow}
\end{equation}
To arrive at a stochastic diffusion equation for the net-baryon density in a rapidly expanding medium we introduce Milne coordinates, i.e.~proper-time $\tau=\sqrt{t^2-z^2}$ and space-time rapidity $y=\frac{1}{2}\ln((t+z)/(t-z))$. In these coordinates Eq.~\eqref{eq:Bjorkenflow} renders to $u^\mu=(\cosh y, 0, 0, \sinh y)$ and $n_B(\tau,y)=\tau n_B(t,z)$.

Using the ansatz Eq.~\eqref{eq:Bjorkenflow} in Eq.~\eqref{eq:current} one obtains a stochastic diffusion equation for the net-baryon density of the form 
\begin{equation}
\label{eq:SDE}
	\partial_{\tau} n_B(\tau, y) = \dfrac{D n_c}{\tau} \partial_y^2 \left(\dfrac{\mu_B(\tau,y)}{T(\tau)}\right) - \partial_y \xi(\tau, y)
	\,.
\end{equation}
The noise field $\xi(\tau,y)=\xi(t,z)$, which is the $y$-component of $\xi^\mu$ in Eq.~\eqref{eq:correlator}, obeys in Milne coordinates the auto-correlation 
\begin{equation}
\label{eq:autocorrelation}
    \langle \xi(\tau, y) \xi(\tau', y') \rangle = \dfrac{2 D n_c }{\tau} \delta(\tau - \tau') \delta(y - y')\,. 
\end{equation}
Equations~\eqref{eq:SDE} and~\eqref{eq:autocorrelation} in Milne coordinates for a constant $D$ may be interpreted as describing a classical stochastic diffusion with a diffusion length, which decreases as a function of time as $1/\tau$. 

In line with $u^\mu$ in Eq.~\eqref{eq:Bjorkenflow}, the temperature cools during the expansion with proper-time as 
\begin{equation}
	T(\tau) = T_i \Big(\frac{\tau_0}{\tau}\Big)^{1/dc_s^2} \,.
	\label{eq:Hubble}
\end{equation}
This describes the temperature evolution starting from an initial temperature $T_i$ at initial proper-time $\tau_0$ similar to the Hubble expansion in cosmology. In this work we consider $d=3$ spatial dimensions and assume a constant squared speed of sound $c_s^2=1/3$ for simplicity. For the numerical results presented in this work we choose $T_i~=~500$~MeV at $\tau_0 = 1$ fm$/$c. 

\subsection{Free energy density functional}
\label{sec:IIB}

In order to solve Eqs.~\eqref{eq:SDE} and~\eqref{eq:autocorrelation} we need to know $\mu_B/T$ as function of $\tau$ and $y$ along a given trajectory in the QCD phase diagram. This can be achieved by making use of either a suitable equilibrium equation of state or by using the thermodynamic relation between $\mu_B$ and the free energy density $F$ reading 
\begin{equation}
 \mu_B = \dfrac{\delta F}{\delta n_B} \,.
 \label{eq:Thermo}
\end{equation}
In this work we pursue the latter approach. 

The equilibrium properties of the medium can be described by a free energy density for an expanding system
\begin{align}
\nonumber
	F[n_B](\tau) = \,& T(\tau)\int {\rm d}y\,{\rm d}^2x_{\perp}   \Bigg(\frac{m^2}{2 \tau n_c^2} n_B(\tau, y)^2 \\ & +\frac{K}{2 \tau^3 n_c^2}\big(\partial_y n_B(\tau,y)\big)^2 
     \,+\frac{\lambda_4}{4 \tau^3 n_c^4}n_B(\tau,y)^4 \Bigg) \,.
\label{eq:GL}
\end{align}
Here $m$ is an effective mass parameter connected with the standard Gaussian diffusion and $K$ is the surface tension parameter related to the kinetic energy of the net-baryon density field. In addition, $F$ includes a term proportional to the coupling strength $\lambda_4$ in order to study non-Gaussian net-baryon density fluctuations arising from the resulting non-linear diffusion equation. The noise term in Eq.~\eqref{eq:SDE} ensures that the equilibrium functional probability distribution is connected with $F$ in the standard way. We note that a similar functional form has been considered in previous studies of the critical dynamics of the net-baryon density, cf.~\cite{Nahrgang:2018afz,Nahrgang:2020yxm}. 

Using the variational relation Eq.~\eqref{eq:Thermo}, we can determine $\mu_B/T$ which expressed as a function of $\tau$ and $y$ reads 
\begin{align}
\nonumber
	\frac{\mu_B}{T}(\tau, y) = \,& \frac{m^2}{\tau n_c^2} n_B(\tau, y) - \frac{K}{n_c^2 \tau^3}\partial_y^2 n_B(\tau, y) \\ & + \frac{\lambda_4}{\tau^3 n_c^4} n_B(\tau,y)^3 \,.
	\label{eq:GLM}
\end{align}
This renders Eq.~\eqref{eq:SDE}, in general, into a non-linear stochastic evolution equation for $n_B(\tau,y)$. The parameters $m$ and $\lambda_4$ are related to the thermodynamic susceptibilities of the net-baryon density using the classical definition of the latter as 
\begin{eqnarray}
\nonumber
	\chi_2 &=& \left(\left.\dfrac{\delta^2 (F/T)}{\delta n_B^2}\right|_{\Delta n_B =0}\right)^{-1} = \dfrac{\tau n_c^2}{m^2} \,, \\
	\chi_4 &=& \left(\left.\dfrac{\delta^4 (F/T)}{\delta n_B^4}\right|_{\Delta n_B =0}\right)^{-1} = \dfrac{\tau^3 n_c^4}{6\lambda_4} \,.
\label{eq:chis}
\end{eqnarray}

In the following, we will study the evolution of net-baryon density fluctuation observables along trajectories of constant $\mu_B$. This implies that the susceptibilities, or the respective parameters, are functions of $T$ at fixed $\mu_B$ and, following Eq.~\eqref{eq:Hubble}, are functions of $\tau$. Injecting Eqs.~\eqref{eq:GLM} and~\eqref{eq:chis} into Eq.~\eqref{eq:SDE} we find 
\begin{equation}
\begin{split}
	\partial_{\tau} n_B = \frac{D n_c}{\tau\chi_2(\tau)} &\partial_y^2 n_B - \frac{D n_c K(\tau)}{\tau} \partial_y^4 n_B \\
	& + \frac{D n_c}{6\,\tau\chi_4(\tau)} \partial_y^2 {n_B}^3 - \partial_y \xi \,.
\end{split}
\label{eq:FullSDE}
\end{equation}
This constitutes an evolution equation directly for the net-baryon density fluctuations $n_B(\tau, y)$, and $K(\tau) = K/\tau^3$ is the Milne equivalent to the surface tension. The prefactors $D n_c/(\tau\chi_2(\tau))$ and $D n_c/(6\tau\chi_4(\tau))$ in the SDE reveal the non-trivial interplay between the diffusion length $D$, the expansion dynamics with increasing proper-time $\tau$ and the equilibrium properties of the medium encoded in the susceptibilities $\chi_{2}$ and $\chi_{4}$. 

\subsection{Parametrization of the susceptibilities in the QCD phase diagram}
\label{sec:IIC}

To model the equilibrium properties of the medium in the QCD phase diagram, we employ parametrizations for the second and fourth order susceptibilities following the procedure advocated in~\cite{Sakaida:2017rtj}. To this end we split the susceptibilities into a regular and a singular part, where the regular part represents all non-critical contributions and the singular part contains the contributions stemming from a conjectured QCD critical point valid only in a specific scaling region around it, via 
\begin{equation}
	\chi_n(T) = \chi_n^{\text{sing}}(T) + \chi_n^{\text{reg}}(T)\,.
	\label{eq:spliting}
\end{equation}
 The singular contributions in $\chi_2$ and $\chi_4$ are, by definition, directly related to the singular contributions in the parameters $m$ and $\lambda_4$ via $\chi_2^{\text{sing}}=\tau n_c^2/m^2_{\text{sing}}$ and $\chi_4^{\text{sing}}=\tau^3 n_c^4/(6\lambda_{4,\text{sing}})$. The latter, as functions of $T$ and $\mu_B$, are obtained from a matching to the susceptibilities of the scaling equation of state of the $3$-dimensional Ising model~\cite{Guida:1996ep} as explained in more detail in~\cite{Bluhm:2016byc}. Correspondingly, $m^2_{\text{sing}}$ and $\lambda_{4,\text{sing}}$ are functions of the correlation length and given as $m^2_{\text{sing}}=(\xi_0\xi^2)^{-1}$ and $\lambda_{4,\text{sing}}= n_c \widetilde\lambda_4(\xi/\xi_0)^{-1}$. As in~\cite{Nahrgang:2018afz,Nahrgang:2020yxm}, we will use constant values for the appearing parameters $\xi_0=0.48$~fm and $\widetilde\lambda_4=10$ throughout this work. This implies, in particular, that $\chi_4$ in this work shows no sign changes across the phase transition~\cite{Bluhm:2016trm}. In this model the critical point is placed at $T_c=150$ MeV and $\mu_{B,c} = 390$ MeV.

The regular part is given by the smooth connection between two limiting values, see~\cite{Sakaida:2017rtj}, for low and high $T$ in the hadronic phase and for the QGP following 
\begin{equation}
	\chi_n^{\text{reg}}(T) = \chi_{0,n}^\text{H} + \big(\chi_{0,n}^\text{QGP} - \chi_{0,n}^\text{H} \big) S(T) \,,
	\label{eq:chi_reg}
\end{equation}
where 
\begin{equation}
    S(T) = \frac{1}{2}\left(1 + \tanh{\left[\frac{T-T_c}{\delta T}\right]}\right) 
    \label{eq:S}
\end{equation}
accomplishes the smooth connection. In Eq.~\eqref{eq:S}, ${\delta T=10}$~MeV is the width of the transition between the limiting values. The parameters $\chi_{0,n}^\text{QGP}$ and $\chi_{0,n}^\text{H}$ are fixed by imposing conditions for the values reached by the full susceptibilities $\chi_n$ in Eq.~\eqref{eq:spliting} at an initial, high temperature and at a final, low $T$. Numerical values for the latter will be discussed in the following. 

\subsection{Connection with susceptibilities from lattice QCD}
\label{sec:IID}

The relations in Eq.~\eqref{eq:chis} define net-baryon number susceptibilities per unit volume expressed in Milne coordinates, i.e. per unit of rapidity and transverse area. For relating them to the lattice equilibrium susceptibilities, we follow the arguments in~\cite{Asakawa:2000wh,Sakaida:2017rtj}. At early times/high temperatures and late times/low temperatures the equilibrium fluctuations per unit of rapidity $\langle {\Delta N_B}^2 \rangle/\Delta y$ are expected to be constant as a function of temperature~\cite{Sakaida:2017rtj}. To obtain the limiting values from lattice QCD results, we use the fact that the entropy per unit rapidity $S/\Delta y$ is conserved as a function of $\tau$ in the Bjorken expansion described by ideal fluid dynamics that may be justified at sufficiently large collision energies.
In this case $\chi_2$ is proportional to 
\begin{equation}
    X_2 = \frac{\langle {\Delta N_B}^2 \rangle}{S} = \frac{\chi_2^{B,{\rm latt}}}{s/T^3} \,,
\end{equation}
during the time evolution. 
The temperature dependence of $X_2$ is obtained from lattice QCD results for $\chi_2^{B,{\rm latt}}$ and $s/T^3$~\cite{Cheng:2008zh,Bazavov:2017dus,Bollweg:2021vqf}, which indeed approaches constants at low and high $T$.
By neglecting the proportionality coefficient $S/\Delta y$, we can therefore at high and low temperature identify $\chi_2(\tau)$ with $X_2$, for which from~\cite{Cheng:2008zh,Bazavov:2017dus} we find as numerical values $X_2^{\text{QGP}} = 0.02$ at $T = 280$~MeV for the QGP and $X_2^{\text{H}} = 0.01$ at $T = 130$~MeV in the hadronic phase. This allows us to determine, in connection with lattice QCD results, values for the parameters $\chi_{0,2}^\text{QGP}$ and $\chi_{0,2}^\text{H}$ needed for the parametrization of $\chi_2(\tau)$ discussed in section~\ref{sec:IIC}. 

\begin{figure}
    \hspace*{-0.4cm}\includegraphics[scale=0.24]{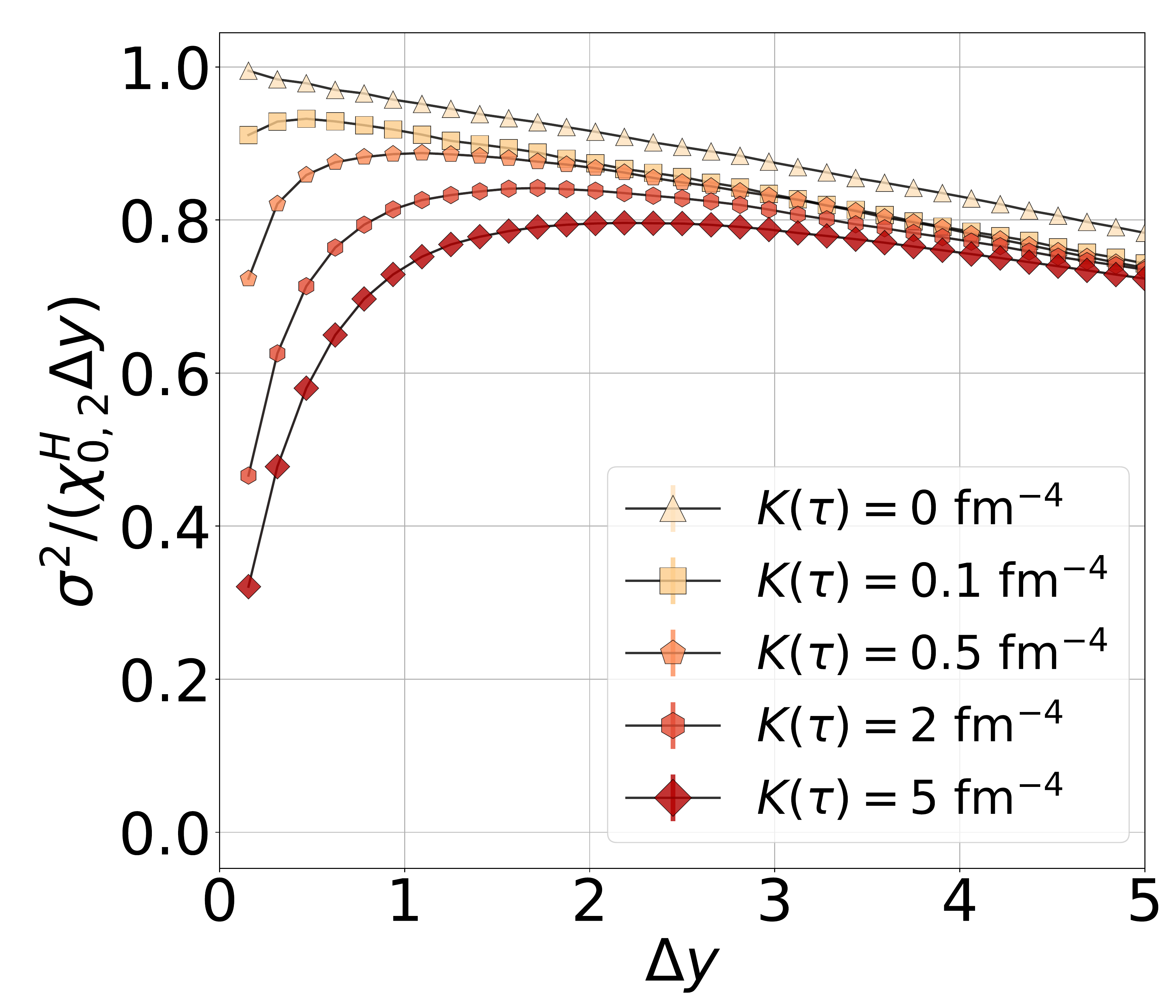}
	\caption{(Color online) The late-time value (at $T = 100$~MeV) of the scaled second-order cumulant of $n_B$, $\sigma^2/(\chi_{0,2}^H \Delta y)$ for a non-critical trajectory as a function of the rapidity window $\Delta y$ for $\lambda_4 = 0$ and several constant values of $K(\tau)$.} 
    \label{fig:Kint}
\end{figure}

We note that a strict identification of these quantities is only possible in the linear Gaussian model ($\lambda_4=K=0$), where the quantitative values can be scaled with the respective $\chi_2^\text{H}$. In this case, $\chi_2(\tau)$ is equal to the equilibrium variance of the net-baryon number at vanishing rapidity window, as discussed in~\cite{Sakaida:2017rtj}, for both the high-temperature QGP and the low-temperature hadronic phase. For $K\to0$ this is verified for our model, shown at late times in the hadronic phase, see Fig~\ref{fig:Kint}. We show in addition the dependence of the second-order cumulant of $n_B$, $\sigma^2$ (as defined below), on the rapidity window for a couple of values of $K(\tau)$. Since in this work we take $K(\tau)=2$~fm$^{-4}$ and look predominantly at $\Delta y = 1$, we expect a reduction of approximately $20\%$ of the final value in the hadronic phase, compared to the input from lattice QCD. 

\begin{figure}
    \hspace*{-0.4cm}\includegraphics[scale=0.24]{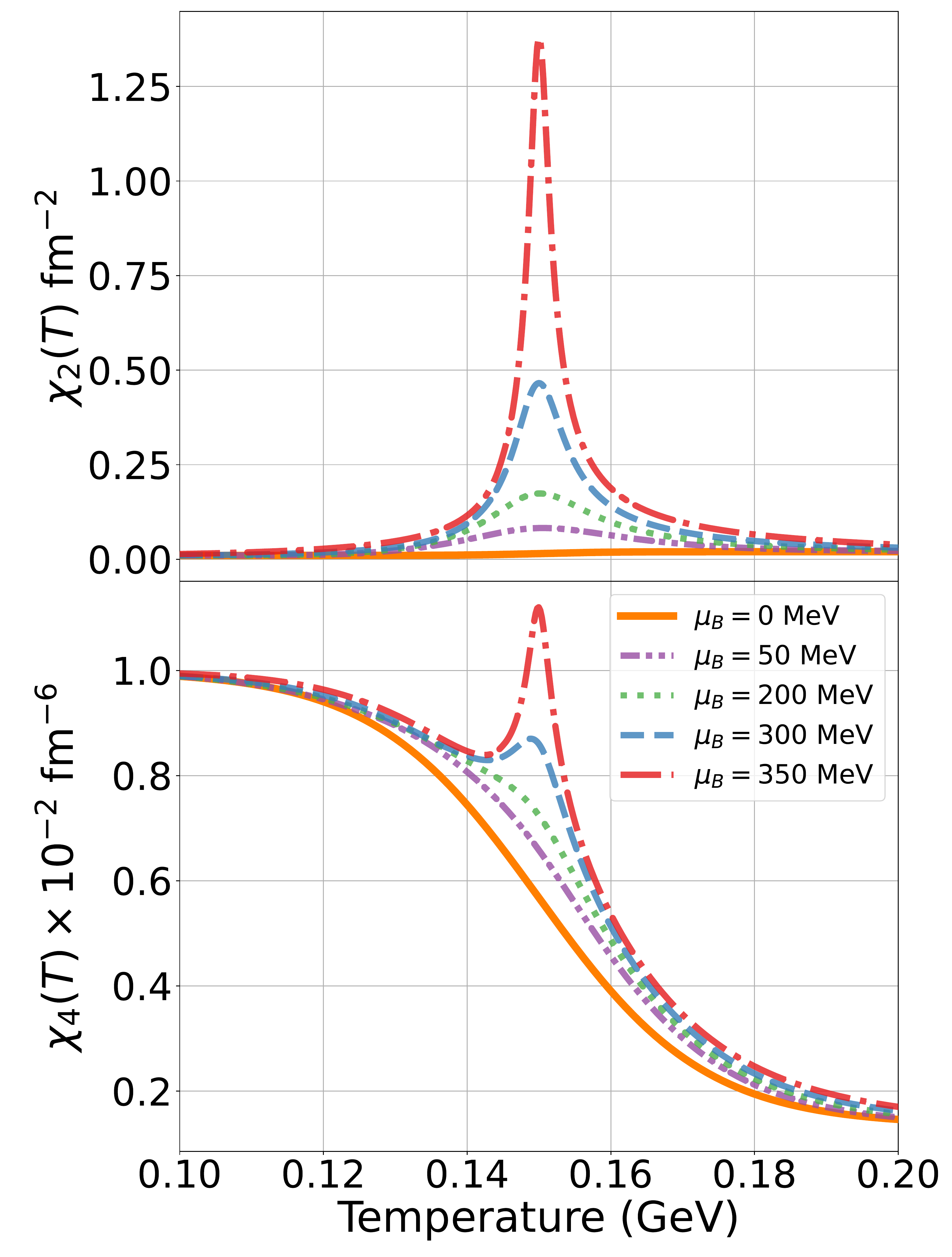}
	\caption{(Color online) The second order (upper panel) and fourth order (lower panel) susceptibilities per unit of rapidity and transverse area as a function of the temperature for different values of the baryo-chemical potential. The values reached in the hadronic phase (at low $T$) and in the QGP (at high $T$) are $\chi_2^{\text{H}}=0.01$ fm$^{-2}$, $\chi_2^{\text{QGP}}=0.02$~fm$^{-2}$, $\chi_4^{\text{H}} = 0.01$ fm$^{-6}$ and $\chi_4^{\text{QGP}} = 0.00135$ fm$^{-6}$.} 
    \label{fig:susceptibility}
\end{figure}

In order to extend the discussion to the fourth order susceptibility per unit of rapidity and transverse area we make use of known relations between $\chi^B_2$ and $\chi^B_4$ in the low and high temperature phases. For low $T$, based on the hadron resonance gas model, one has $\chi^B_4 = \chi^B_2$. For high $T$, by considering a free gas of quarks and gluons, one finds $\chi^B_4 =2/(3 \pi^2) \chi^B_2$. These relations are extended to determine $X_4$ in connection with $X_2$ by dividing $\chi^B_{2,4}$ with $s/T^3$. In line with recent lattice QCD results \cite{Cheng:2008zh,Bazavov:2017dus} we use
$X_4^{\text{H}} = X_2^{\text{H}}$ in the hadronic phase and $X_4^{\text{QGP}} = 2/(3 \pi^2) X_2^{\text{QGP}}$ for the QGP as limiting values. From these, values for the parameters $\chi_{0,4}^\text{QGP}$ and $\chi_{0,4}^\text{H}$ entering the parametrization of $\chi_4$ are obtained. In Fig.~\ref{fig:susceptibility}, we show the resulting parametrizations of the second and fourth order susceptibilities per unit of rapidity and transverse area, $\chi_2$ and $\chi_4$, as functions of $T$ for different constant $\mu_B$. We note that for this work we ignore a possible $\mu_B$ dependence of $\chi_{2(4)}^\text{QGP}$ and $\chi_{2(4)}^\text{H}$.

\section{Two-point fluctuation observables in the Gaussian approximation \label{sec:III}}

In this section, the numerical results for simulations of the SDE in the Gaussian approximation are shown to reproduce analytic expectations which can be derived in this limit. This is an essential step, as it confirms the ability of the numerics to simulate the expected equilibrium power spectrum of the fluctuations in the linear approximation. For the numerics discussed in this section we choose $n_B(\tau_0,y)=0$ as initial condition for the net-baryon density fluctuations and $D=1$~fm.\\
We impose $\lambda_4 = 0$ in the free energy density functional in Eq.~\eqref{eq:GL}. The SDE in Eq.~\eqref{eq:FullSDE} then takes the following form  
\begin{equation}
	\partial_{\tau} n_B = \frac{D n_c}{\tau\chi_2(\tau)} \partial_y^2 n_B 
	- \frac{D n_c K(\tau)}{\tau} \partial_y^4 n_B 
	- \partial_y \xi \,.
\label{eq:GS}
\end{equation}
Since Eq.~\eqref{eq:GS} is linear in the net-baryon density fluctuations $n_B(\tau,y)$, we can obtain a formal solution for the spatial Fourier transform of the fluctuations as 
\begin{align}
\nonumber
 n_B(\tau,q) & = \,n_B(\tau_0,q)\,e^{-q^2 d(\tau_0, \tau)^2/2 -q^4 k(\tau_0, \tau)^4/4} \\
 & + \int_{\tau_0}^{\tau} i q \,\xi(\tau',q) \,e^{- q^2 d(\tau', \tau)^2/2 - q^4 k(\tau', \tau)^4/4} \, \mathrm d\tau' \,,
\label{eq:FormSol}
\end{align}
where $\tau_0$ is the initial proper time. In Eq.~\eqref{eq:FormSol}, the time-integrated diffusion and surface tension coefficients for the expanding medium between two proper times ${\tau_1 < \tau_2}$, $d(\tau_1, \tau_2)$ and $k(\tau_1, \tau_2)$, appear and are defined as 
\begin{align}
\label{eq:DK1}
 d(\tau_1, \tau_2)^2 = & \, 2D n_c \int_{\tau_1}^{\tau_2} \frac{1}{\tau'\chi_2(\tau')} \rm d\tau' \,,
 \\
 k(\tau_1, \tau_2)^4 = & \,4 D n_c\int_{\tau_1}^{\tau_2} \frac{K(\tau)}{\tau'} \rm d\tau' \,.
\label{eq:DK2}
\end{align}

The solution for $n_B(\tau,q)$, Eq.~\eqref{eq:FormSol}, allows us to analytically study the evolution of fluctuations in the continuum limit. In the Gaussian approximation, only two-point fluctuation observables are relevant for which we present the structure factor and its inverse Fourier transform, the two-point correlation function, at equal proper-time in the following. 

\subsection{Structure factor at equal proper-time}
\label{sec:IIIA}

From the continuum solution in Eq.~\eqref{eq:FormSol} we may construct the two-point correlation function in Fourier space at equal proper-time as $G(\tau, q_1,q_2)=\langle n_B(\tau,q_1) \, n_B(\tau,q_2)\rangle$. For a translationally invariant system all information is contained in the structure factor which, because $n_B(\tau,y)$ is real, is related to the two-point correlation function with $q_1 = - q_2$ via 
\begin{equation}
 S(\tau,q_1)\delta(q_1+q_2) = G(\tau,q_1,q_2) \,.
\label{eq:FormSK}
\end{equation}

Assuming that there are no correlations between different modes $q_i$ and $q_j$ of the noise field for all proper times $\tau$, i.e.~$\langle \xi(\tau,q_i)\,\xi(\tau,q_j)\rangle = 0$, and between the initial net-baryon density fluctuations and the noise for all $\tau\geq\tau_0$, i.e.~$\langle n_B(\tau_0,q_i)\,\xi(\tau,q_j)\rangle = 0$, we find the following structure factor in the Gaussian approximation 
\begin{align}
\nonumber
	S(\tau,q) & = \,S(\tau_0,q)\,e^{\alpha(q) d(\tau_0, \tau)^2 + \beta(q) k(\tau_0, \tau)^4/2} \\
	& \,\,\,\,+ \gamma(q) \int_{\tau_0}^{\tau} \frac{\rm d \tau'}{\tau'} e^{\alpha(q) d(\tau', \tau)^2 + \beta(q) k(\tau', \tau)^4/2}  \,,
\label{eq:Sk}
\end{align}
where 
\begin{align}
\nonumber
 \alpha(q) & = -q^2\,,\\
 \beta(q) & = -q^4\label{eq:CoeffCont}\,,\\
\nonumber
\gamma(q) & = 2 D n_c q^2\,.
\end{align}
We assume that the initial net-baryon density fluctuations are local in space with a variance $\chi_0$, i.e.~$\langle n_B(\tau_0,y_1)\, n_B(\tau_0,y_2)\rangle = \chi_0\delta(y_1-y_2)$. Then, $S(\tau_0,q)$ is the inverse Fourier transform of this initial correlation function. 

A direct comparison between Eqs.~(\ref{eq:Sk}~-~\ref{eq:CoeffCont}) and our numerical calculations is not possible due to the lattice spacing dependence of the amplitude of the noise in the discretized space-time. Therefore, we derive a formal solution for $n_B$ in discretized space for the Gaussian limit and find the associated structure factor at finite lattice spacing $\delta y$, see Appendix~\ref{sec:AppA} for more details. This discretized solution depends on the particular discretization scheme used for the numerical integration. We find $S(\tau,q)$ to be formally the same as in Eq.~\eqref{eq:Sk} but the discretized counterparts of the coefficients in Eq.~\eqref{eq:CoeffCont} now read 
\begin{align}
\nonumber
\alpha(q) & = - \frac{2(1-\cos(\Delta q))}{{\delta y}^2}\,,\\
\beta(q) & = - \frac{4(1-\cos(\Delta q))^2}{{\delta y}^4}\label{eq:CoeffDis}\,,\\
\nonumber
\gamma(q) & = \frac{8 D n_c}{ {\delta y}^2} \sin(\Delta q/2)^2\,.
\end{align}
Here $\Delta q = q \,\delta y$ for a cell-size in space-time rapidity $\delta y$ defined as $\delta y = L/N$, where $N$ is the number of cells in the $y$-direction within $L$ units of rapidity. Correspondingly, only a finite number of discrete modes with wave-vector length $q$ are realized. 

We note that in the limit $\Delta q \rightarrow 0$ the coefficients in Eq.~\eqref{eq:CoeffDis} tend to those in Eq.~\eqref{eq:CoeffCont}. Therefore, the structure factor in discretized space approaches its continuum counterpart either for small $q$ at a given $\delta y$ or with increasing resolution $\delta y \rightarrow 0$. 

\begin{figure*}
 \includegraphics[scale=0.2]{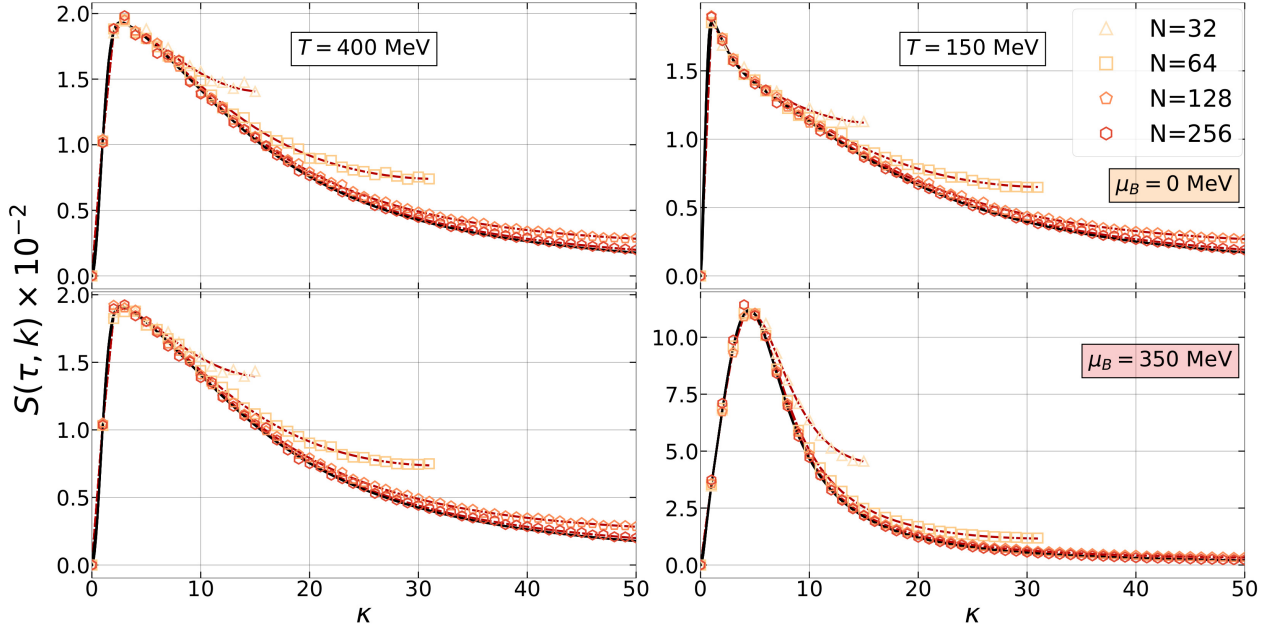}
	\caption{(Color online) The structure factor $S(\tau,\kappa)$ of the net-baryon density fluctuations at $T = 400$~MeV (left panels) and at $T = T_c = 150$~MeV (right panels) as a function of the wave-number $\kappa = Lq/(2 \pi)$. The continuum solution (solid line) is compared to our numerical results (symbols) and the solutions in discretized space (dashed lines) for different numbers of cells $N=32, 64, 128, 256$ within $L=20$ units of rapidity. In the upper panels the comparison is shown for a trajectory at constant $\mu_B = 0$~MeV far away from the critical point and in the lower panels for a trajectory with $\mu_B = 350$~MeV passing near the critical point. 
   }
 \label{fig:Sk150}
\end{figure*}
A comparison between the continuum solution (solid line), the results of our numerical calculations (symbols) and the discretized solutions (dashed lines) for different resolutions $\delta y$ at two different $T$ and two different $\mu_B$ is presented in Fig.~\ref{fig:Sk150}. Here, we vary $N$ while keeping $L=20$ fixed. Because of the symmetry property of the discretized structure factor $S(\tau,\kappa)=S(\tau,N-\kappa)$ we show the results only for $0\le\kappa\le N/2$. The wave-number $\kappa = Lq/(2 \pi)$ ranging from $0$ to $N$ is used to have a coherent variable for continuum, discretized and numerical calculations. For the chosen initial condition we have $S(\tau_0,q) = 0$ and Eq.~\eqref{eq:Sk} simplifies significantly. We observe that the numerical results perfectly reproduce the analytic expectations in discretized space for all $T$ and $\mu_B$ and resolutions. Moreover, as $N$ is increased the structure factor approaches more and more the continuum result. For $N=128$ we find already a reasonable agreement between the numerics and the continuum for small and intermediate wave-numbers. For this reason and to optimize the computational effort, we choose $N=128$ in the remainder of this work. 

In Fig.~\ref{fig:Sk150}, we observe that far away from the critical point, at $\mu_B=0$~MeV, the power spectrum of the fluctuations remains almost unchanged when approaching the pseudo-critical temperature as expected from the behavior of $\chi_2$. For the trajectory at $\mu_B=350$~MeV, which passes near the critical point, one finds the expected significant enhancement in the amplitude of the fluctuations at $T=T_c$ (note the different scale of the y-axis). Moreover, we observe a slight shift in the position of the maximum of $S(\tau,\kappa)$ toward larger wave-numbers $\kappa$, i.e.~smaller wave-lengths. In contrast, far outside the critical region for $T=400$~MeV the result resembles the situation on the non-critical trajectory.

\subsection{Correlation function at equal proper-time}
\label{sec:IIIB}

The two-point correlation function of the  net-baryon density fluctuations in the Gaussian approximation can be obtained by calculating the inverse Fourier transform of the structure factor in Eq.~\eqref{eq:Sk}. Its behavior as a function of proper-time and space-time rapidity is strongly affected by the size of the system and the total charge conservation, the diffusion length and the expansion rate. 

\begin{figure}
 \includegraphics[scale=0.21]{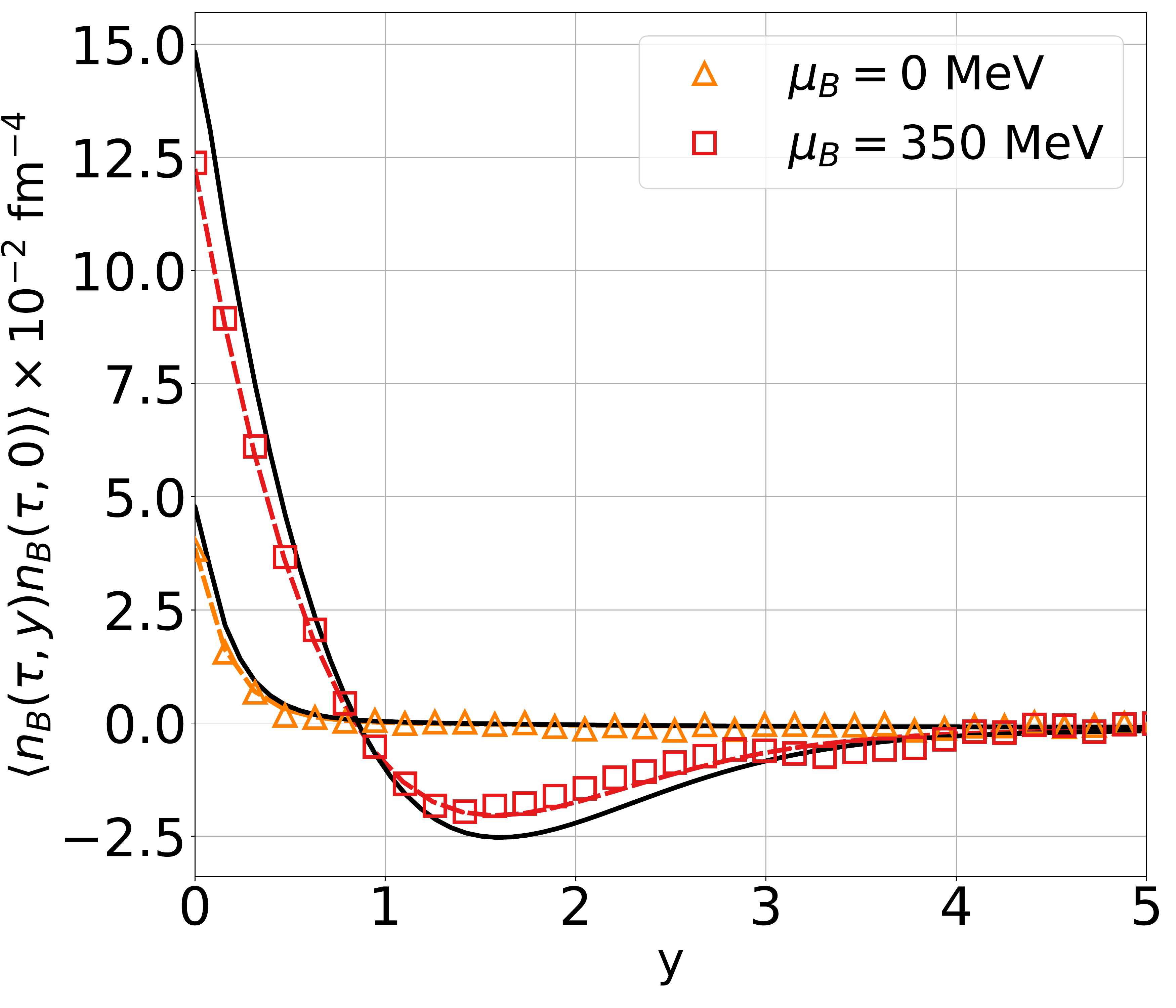}
	\caption{(Color online) The correlation function at equal proper-time in the Gaussian approximation as a function of space-time rapidity for different trajectories with constant $\mu_B$ at $T = 150$~MeV. The symbols represent our numerical results, and the solid and dashed lines are the analytic solutions obtained through inverse Fourier transformation of the continuum and the discretized structure factor, respectively.}
 \label{fig:CorrGS}
\end{figure}
In Fig.~\ref{fig:CorrGS}, the correlation function for the Gaussian limit is presented at $T = T_c$ for a critical and a non-critical trajectory. In contrast to the analytic expectation for infinite and static systems, the correlation function for a finite and rapidly expanding system becomes negative at a certain distance in rapidity. This reflects the conservation of the net-baryon number, since in a finite system the integral of the correlation function over the entire system must vanish. In an expanding medium large fluctuations are balanced locally by an anti-correlation and the correlation function goes back to zero for large distances. It is expected that with increasing $D$ this anti-correlation is diffused over the entire finite system, as seen in~\cite{Nahrgang:2020yxm}. In Fig.~\ref{fig:CorrGS} one clearly observes that as one approaches the critical point with increasing $\mu_B$, correlations increase significantly at small distances and, as a consequence, the anti-correlations are also enhanced and remain visible over a rather large region in space-time rapidity up to $y\approx 1-2.5$ for the considered $D=1$~fm.

Note that in a dynamical system the resolution dependence of the correlation function is not limited to small rapidities, as the corresponding change in the anti-correlations is not diffused over the entire system. Therefore, we observe in Fig.~\ref{fig:CorrGS} a slight difference between the discretized (numerical and analytical) results and the continuum result.

The presented studies of the resolution dependence of the structure factor and of the impact of total charge conservation on the correlation function in a finite-size system serve as successful benchmark tests for our numerical approach. In addition, we observe the basic effect of an increase of the fluctuations as the critical point is approached with increasing $\mu_B$. This motivates us to proceed and include as a next step the non-linear coupling term, see Eq.~\eqref{eq:FullSDE}. In this case, no analytic expressions are available. 

\section{Dynamics of fluctuation observables}
\label{sec:Dynamics}

In this section, we apply the full SDE, as given in Eq.~\eqref{eq:FullSDE}, in order to study the dynamics of the higher-order cumulants of the net-baryon density fluctuations. We then investigate the dependence on the diffusion length $D$ and the considered rapidity window $\Delta y$. For the numerics discussed in this section we choose equilibrated initial conditions, i.e. a system where the fluctuations are initially characterized by a (local) equilibrium variance and an equilibrium correlation function. This implies that before we let the system rapidly expand and cool from $T_i=500$~MeV at $\tau_0=1$fm$/$c, we perform a significant amount of numerical evolution steps at fixed $T_i$ and $\tau_0$. In this way we ensure that the studied fluctuation observables start from a constant, evolution-step independent value at $\tau_0$.

We calculate the second and fourth order cumulants of the net-baryon density distribution in a given rapidity window $\Delta y$. They are related to the variance and the kurtosis of these distributions. To compute these quantities, the integral over $\Delta y$ is calculated symmetrically for each noise configuration~$i$ 
\begin{equation}
	n_{\Delta y, i}(\tau) = \int_{-\Delta y/2}^{\Delta y/2} n_{B,i}(\tau, y) {\rm d}y\,,
	\label{eq:aver}
\end{equation}
where $n_{B,i}$ refers to the field of net-baryon density fluctuations for the noise configuration $i$. From the distribution of $n_{\Delta y, i}(\tau)$ over the noise configurations the $n$th centered moments can be calculated as 
\begin{equation}
	m_{n, \Delta y}(\tau) = \frac{1}{N_{\text{conf}}}\sum_{i=1}^{N_{\text{conf}}} \big(n_{\Delta y,i}(\tau) - \langle n_B \rangle \big)^n 
	\label{eq:mom}
\end{equation}
with $\langle n_B \rangle$ the mean value of the distribution of $n_{\Delta y, i}(\tau)$ over the different noise configurations and the sum runs over the number of noise configurations, $N_{\text{conf}}$. The cumulants $\kappa_{n, \Delta y}$ are then defined at order $2$ and $4$ from the moments using the following formula 
\begin{equation}
 \kappa_{2, \Delta y}(\tau) = m_{2, \Delta y} \,, \quad 
 \kappa_{4, \Delta y}(\tau) = m_{4, \Delta y} - 3 m_{2, \Delta y} \,.
 \label{eq:cumulantrelation}
\end{equation}

In particular, the second and the fourth order cumulants, $\kappa_2$ and $\kappa_4$, are related to the variance $\sigma^2$ and the kurtosis $\kappa$ via the relations 
\begin{equation}
	\sigma^2 = \kappa_{2, \Delta y}, \;\;\;\;\;\;\;\; \kappa = \frac{\kappa_{4, \Delta y}}{\kappa_{2, \Delta y}^2}\,.
	\label{eq:varkurt}
\end{equation} 

By looking at integrated quantities such as defined in Eq.~(\ref{eq:aver}), we avoid any residual lattice spacing dependence in a one-dimensional system like the longitudinal expansion studied here. It also allows us to study the dependence of the fluctuation observables on the rapidity window $\Delta y$. 

\subsection{Time evolution of fluctuation observables}

\begin{figure}
\includegraphics[scale=0.26]{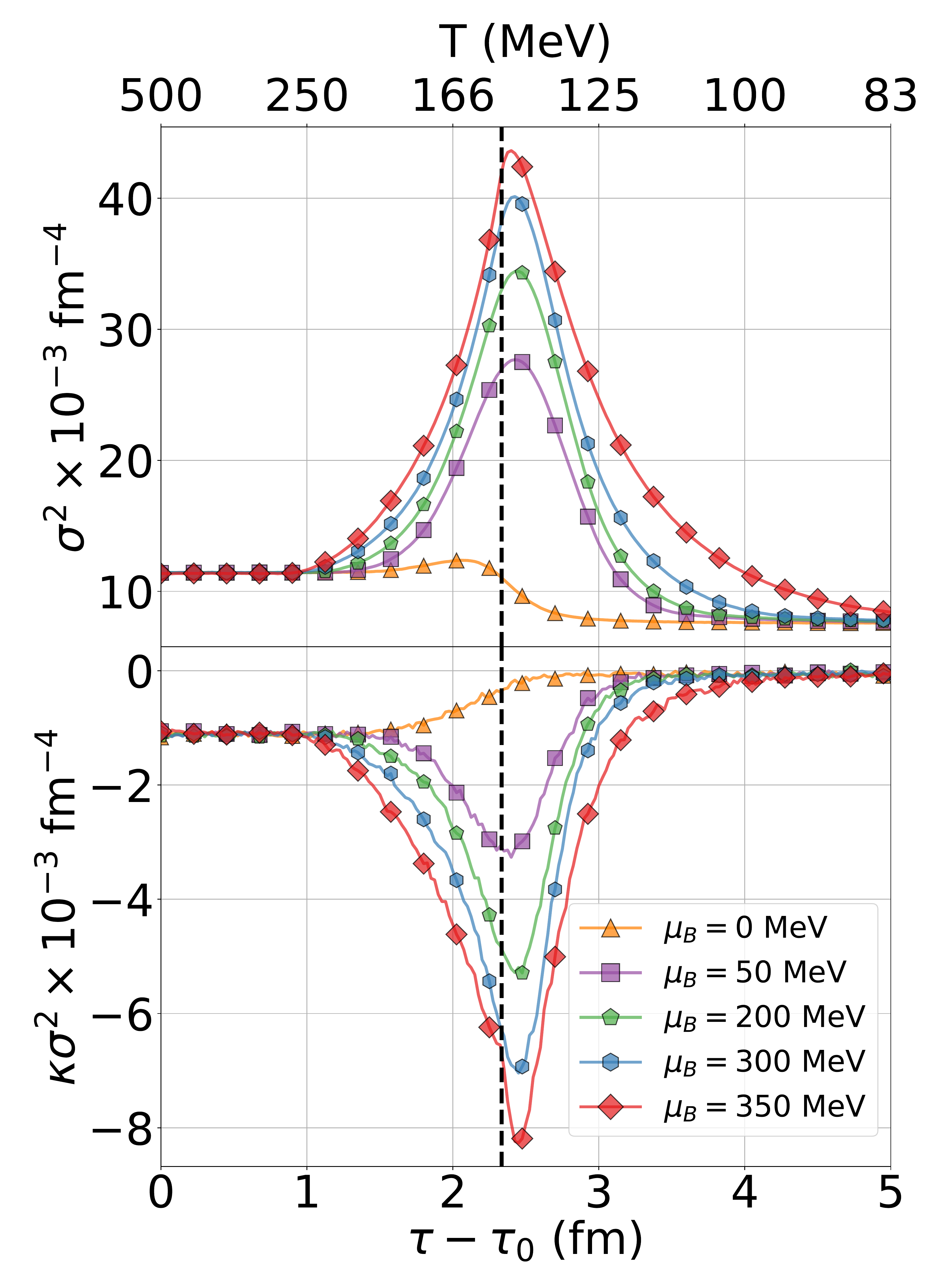}
	\caption{(Color online) The second (upper panel) and fourth (lower panel) order cumulants of $n_B$ within a space-time rapidity window of $\Delta y = 1$ as a function of proper-time for different constant $\mu_B$ (the corresponding temperature is displayed on the upper x-axis). The vertical dashed-line indicates the proper-time where the pseudo-critical temperature is reached. Here, the diffusion length is set to $D=1$~fm. } 
 \label{fig:timeVarKurt}
\end{figure}
\begin{figure}
 \includegraphics[scale=0.23]{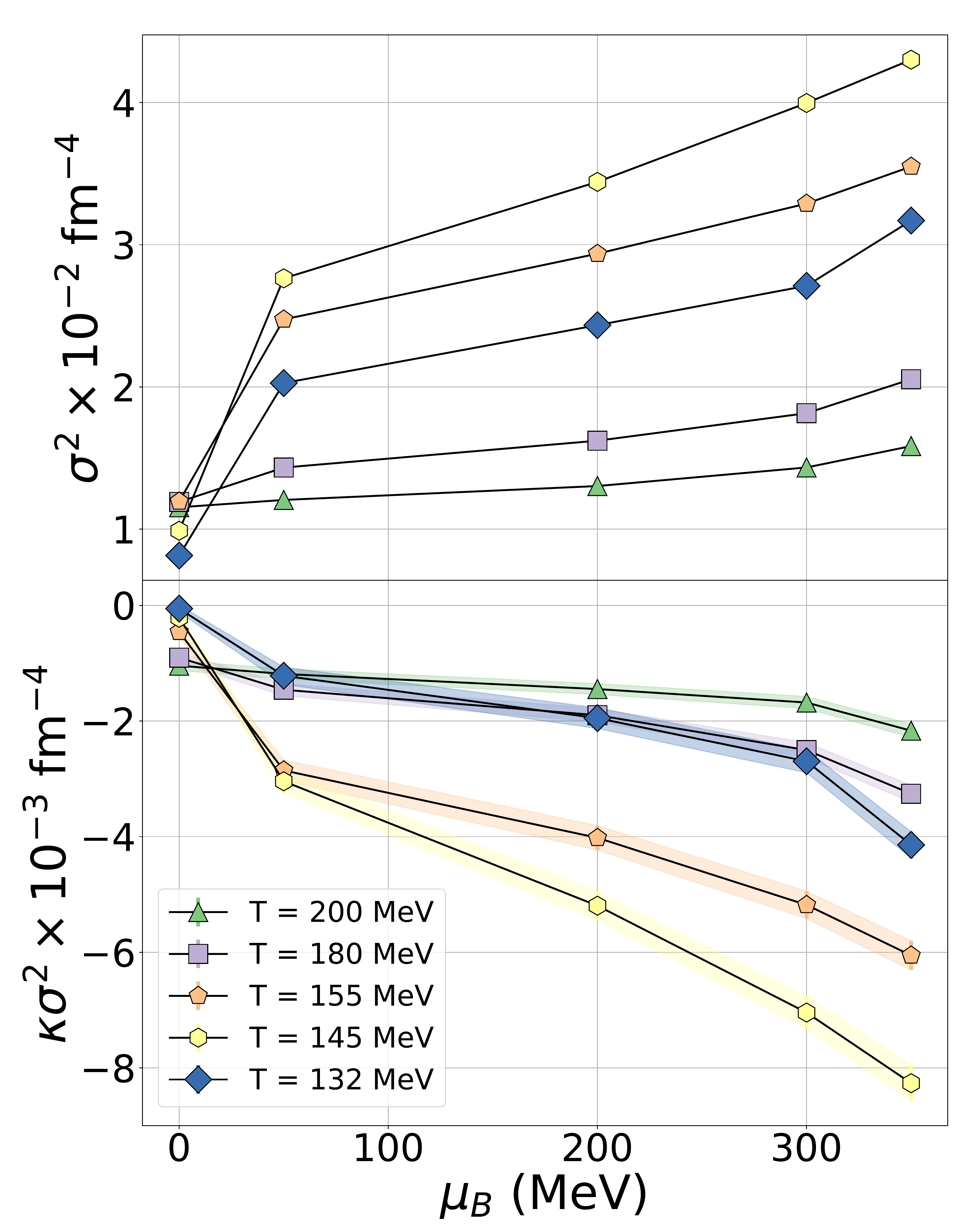}
	\caption{(Color online) The second (upper panel) and fourth (lower panel) order cumulants of $n_B$ within rapidity window $\Delta y = 1$ as a function of $\mu_B$ for different fixed $T$ and $D=1$~fm. The colored bands highlight the statistical uncertainties in the cumulants for a total number of $2.25\times10^6$ noise configurations.} 
 \label{fig:MubVarKurt}
\end{figure}
It is instructive to study the time evolution of the fluctuation observables in order to understand their behavior in the region of the phase diagram where the chemical freeze-out occurs. In Fig.~\ref{fig:timeVarKurt}, the second and fourth order integrated cumulants of the net-baryon density averaged over $2.25\times10^6$ noise configurations are presented for different values of $\mu_B$. The non-critical trajectory $\mu_B=0$~MeV accounts for the behavior of the background fluctuations (solid lines in Fig.~\ref{fig:susceptibility}) namely a smooth connection between the QGP and the hadronic values. As expected from the discussion of Fig.~\ref{fig:Kint}, we observe a reduction of about $20\%$ in $\sigma^2$ at low $T$ compared to the second-order susceptibility for a finite value of $K(\tau)$. 

The fluctuations are largely affected by the presence of a critical point (see Fig.~\ref{fig:timeVarKurt}) despite the rapid expansion in the longitudinal direction. More precisely, after the susceptibilities reach their peak values at $T=150$~MeV (dashed vertical line), the variance and the kurtosis show also a peak value which increases when $\mu_B$ approaches the critical baryo-chemical potential $\mu_{B,c} = 390$ MeV. Moreover, in comparison to the susceptibilities (see Fig.~\ref{fig:susceptibility}) the critical signals are clearly visible in a broadened temperature region. The kurtosis is negative which for positive $\lambda_4$ (and in the absence of a cubic coupling) can already be anticipated from the leading-order term in~\cite{Stephanov:2008qz}.

\begin{figure*}
 \includegraphics[scale=0.25]{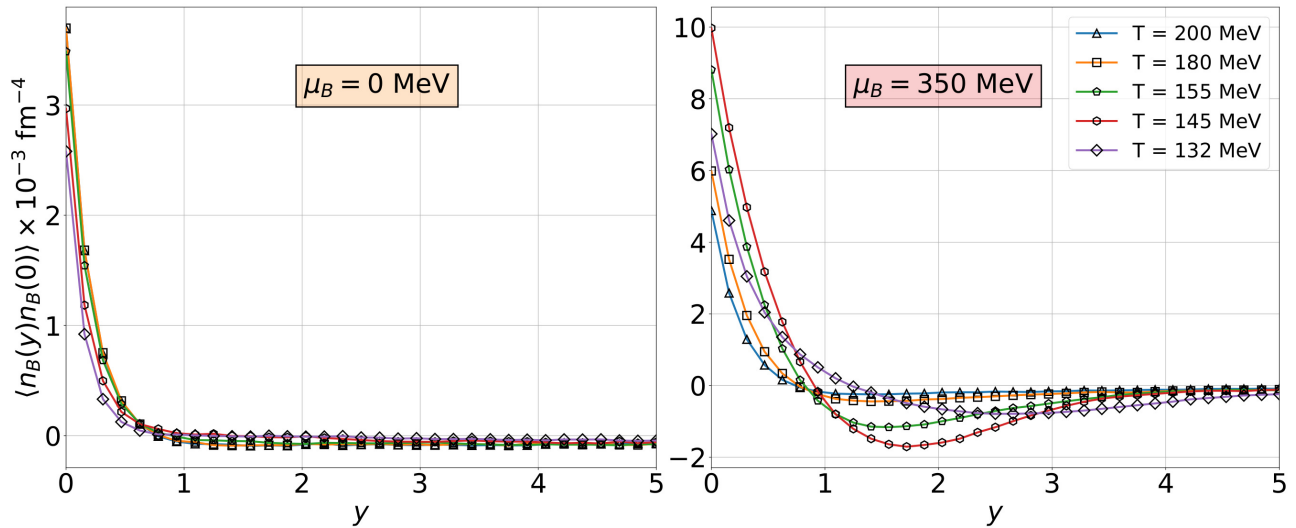}
	\caption{(Color online) The correlation function of $n_B$ as a function of the spatial rapidity $y$ for different fixed $T$ and $D=1$ fm.}
 \label{fig:CorrT}
\end{figure*}
To probe the sensitivity of the observed signal on the freeze-out conditions, an alternative point of view is presented in Fig.~\ref{fig:MubVarKurt}. The integrated second and fourth order cumulants are shown as a function of $\mu_B$ for different temperatures. Consistently with the result in Fig.~\ref{fig:timeVarKurt} a non-monotonic behavior is observed  for the critical trajectories ($\mu_B > 0$~MeV) on a range of $\Delta T \sim 20$~MeV around $T = 150$~MeV corresponding to a time interval of $\Delta \tau \sim 0.6$ fm. 

Finally, in Fig.~\ref{fig:CorrT} we look at the correlation function for the reference trajectory without criticality at $\mu_B=0$~MeV and the most critical trajectory at $\mu_B=350$~MeV for different temperatures. As in Fig.~\ref{fig:CorrGS}, we observe an enhancement of the positive correlations within a range of one unit of rapidity for the critical trajectory near $T_c$. Beyond this an anti-correlation is observed at intermediate rapidities due to global charge conservation. Experimentally, it might be interesting to not only look for the strong enhancement of net-baryon fluctuations in a small rapidity window, but also for the strong anti-correlation of baryons at intermediate rapidity.

\subsection{Diffusion length dependence of fluctuation observables}

\begin{figure}
 \includegraphics[scale=0.23]{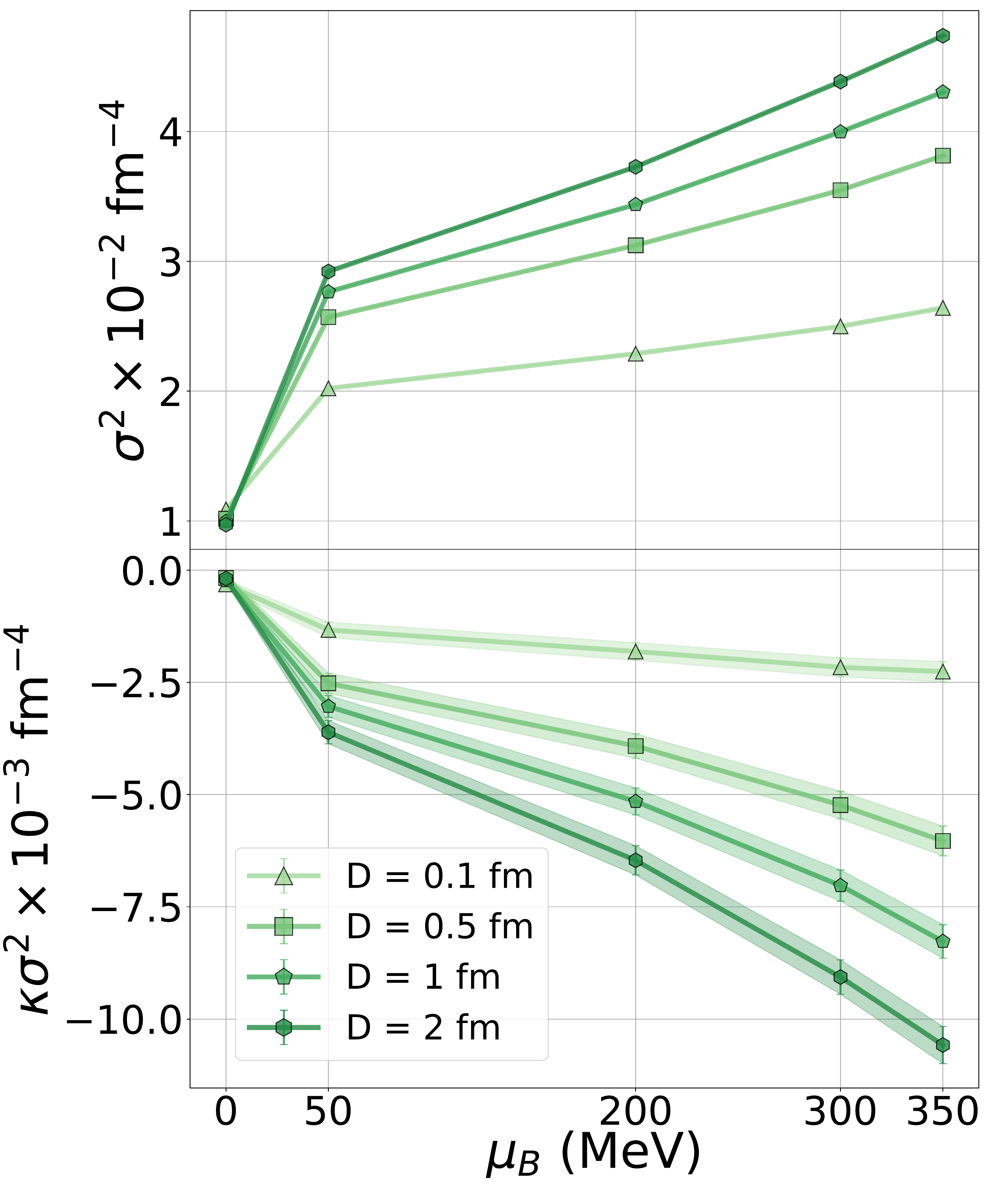}
	\caption{(Color online) The second (upper panel) and fourth (lower panel) order cumulants of $n_B$ within rapidity window $\Delta y = 1$ as a function of $\mu_B$ for different values of the  diffusion length $D$ at fixed $T=145$~MeV. The colored bands correspond to the statistical uncertainties in the cumulants for $2.25\times10^6$ noise configurations.} 
 \label{fig:DVarKurt}
\end{figure}

The diffusion length $D$ in Eq.~\eqref{eq:FullSDE} is inversely related to the relaxation time $\tau_r$ of the fluctuations, see~\cite{Nahrgang:2020yxm}. In our study the diffusion of the fluctuations competes with the rapid expansion of the system. It can be expected that for small diffusion lengths $D$ the system cannot reach equilibrium because the expansion occurs much faster. In the opposite limit of large diffusion lengths $D$ the fluctuation observables should qualitatively resemble those studied in static systems, cf. \cite{Nahrgang:2018afz, Nahrgang:2020yxm}. In Fig.~\ref{fig:DVarKurt}, we show the dependence of the cumulants on the diffusion length and thus on the relaxation time. At fixed (assumed) freeze-out temperature $T_f = 145$ MeV the strength of the fluctuation observables in the rapidity window $\Delta y = 1$ increases for all considered trajectories when we increase the diffusion length.

\begin{figure}
 \includegraphics[scale=0.23]{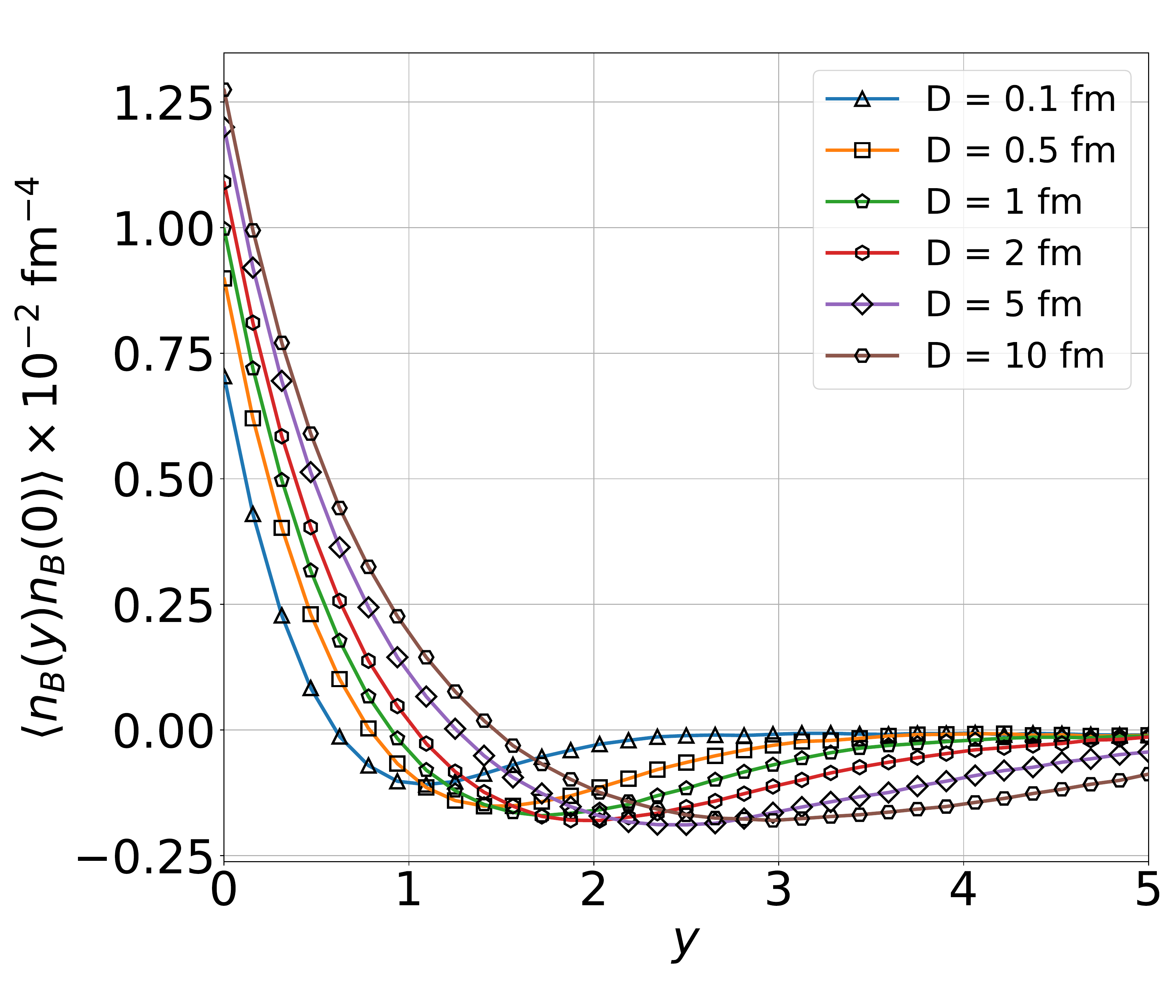}
	\caption{(Color online) The correlation function of $n_B$ as a function of the spatial rapidity $y$ for different $D$ at $\mu_B = 350$~MeV and $T = 145$~MeV.} 
 \label{fig:CorrD}
\end{figure}
For very small diffusion lengths it is found that the critical signals at freeze-out are only slightly increasing when approaching the critical point with increasing $\mu_B$. In this case the rapid expansion of the system has almost completely washed out the critical signal. When we look at the correlation function for different diffusion lengths $D$ in Fig.~\ref{fig:CorrD}, we clearly observe the transition from expansion-dominated, i.e. small $D$, to diffusion-dominated, i.e. large $D$, behavior. For small $D$ the anti-correlation that develops due to charge conservation is trapped at smaller rapidity, while with increasing $D$ it diffuses more and more into the entire system and the correlation function starts to resemble the one we know from static systems~\cite{Nahrgang:2018afz, Nahrgang:2020yxm}. This study demonstrates that the precise determination of this parameter is important for obtaining quantitative results.

\subsection{Rapidity dependence of fluctuation observables at freeze-out}

In the experimental situation, the fluctuations will always be measured in a certain rapidity window and be limited by the total rapidity acceptance of the dectector. Using the definition of the integrated quantities defined via Eq.~(\ref{eq:aver}), we can study the fluctuation observables as a function of the rapidity window at the (assumed) freeze-out temperature $T_f = 145$ MeV.

\begin{figure}
 \includegraphics[scale=0.23]{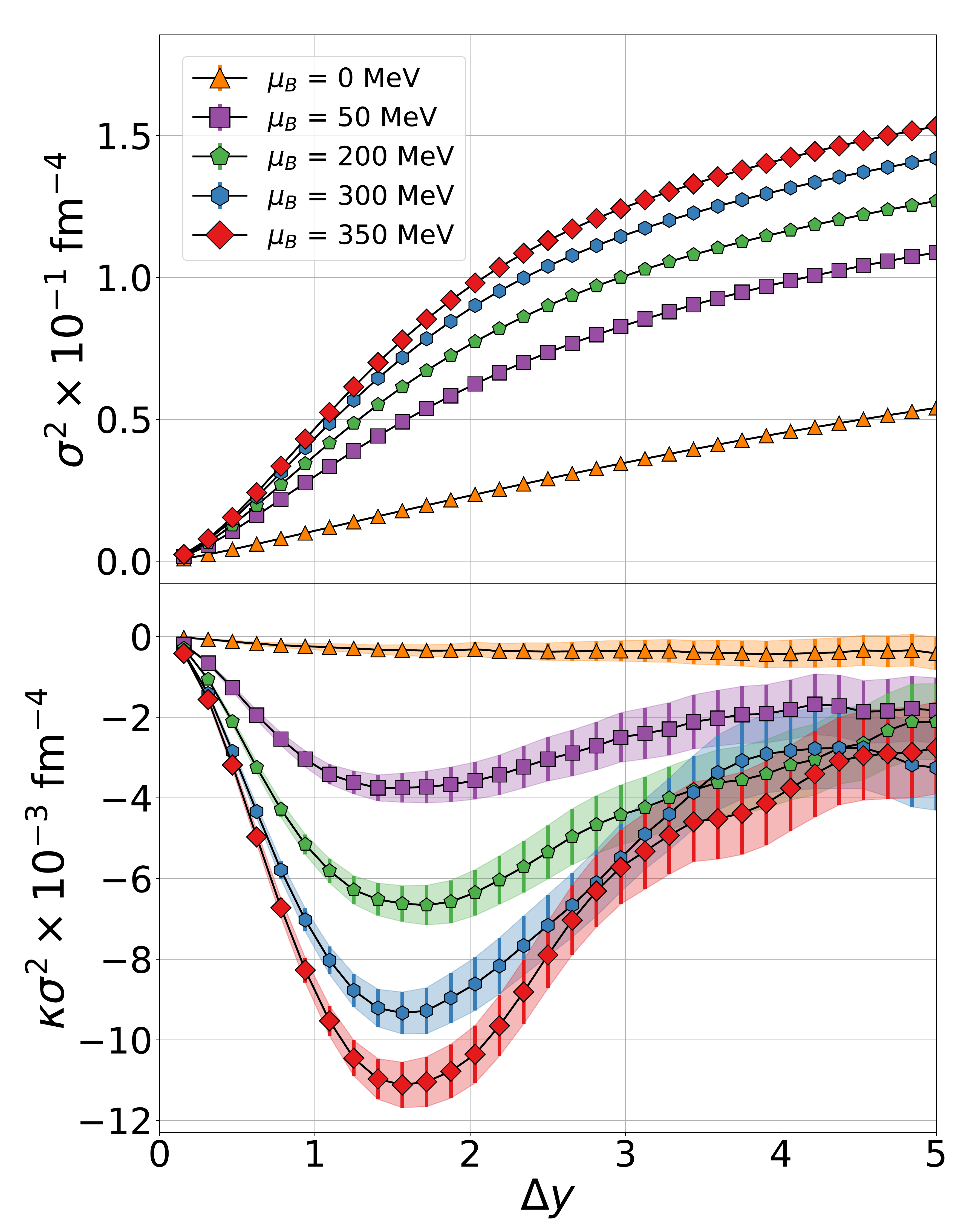}
	\caption{(Color online) The second (upper panel) and fourth (lower panel) order integrated cumulants of $n_B$ as a function of the rapidity window $\Delta y$ at $T=145$~MeV for different  trajectories at constant $\mu_B$. Here, the diffusion length is set to $D=1$~fm. The error bands indicate statistical uncertainties for $2.25 \times10^6$ noise configurations.} 
 \label{fig:DYVarKurt}
\end{figure}
In Fig.~\ref{fig:DYVarKurt}, we compare the second and fourth order integrated cumulants of the net-baryon density fluctuations as defined in Eqs.~(\ref{eq:mom})~-~(\ref{eq:varkurt}) at $T_f$ for several values of $\mu_B$. A monotonic increase of the second-order cumulant (upper panel) can be observed for all trajectories. This second-order cumulant corresponds to the double integral over the point-wise correlation function and can be expressed as \cite{Sakaida:2017rtj}
\begin{align}
   \sigma^2 & = \int_{-\Delta y/2}^{\Delta y/2}\int_{-\Delta y/2}^{\Delta y/2} \langle n_{B}(\tau, y_1) n_{B}(\tau, y_2)\rangle{\rm d}y_1{\rm d}y_2,\nonumber \\
            & = \int_{-\Delta y}^{\Delta y}\left(\Delta y - |\bar{y}|\right)\langle n_{B}(\tau, \bar{y}) n_{B}(\tau, 0)\rangle{\rm d}\bar{y}\,,
\end{align}
where the correlation function only depends on the difference $\bar{y}=y_1-y_2$. Since the simple integral over the correlation function in terms of $\bar{y}$ is zero when the integration area approaches half of the system size (due to charge conservation and symmetry) the double integral plateaus at the same point.

The fourth-order cumulant presents a qualitative change in its shape between the non-critical trajectory at $\mu_B=0$~MeV and the trajectories impacted by the critical point. For the non-critical trajectory, the rapidity window dependence is monotonically decreasing and for critical trajectories it is non-monotonic, with a pronounced minimum. This non-monotonic behavior of the fourth-order cumulant survives the rapid expansion of the system for a diffusion length $D=1$~fm and consequently, if observed experimentally, is a strong indication for the presence of the critical point. 

\begin{figure}
 \includegraphics[scale=0.23]{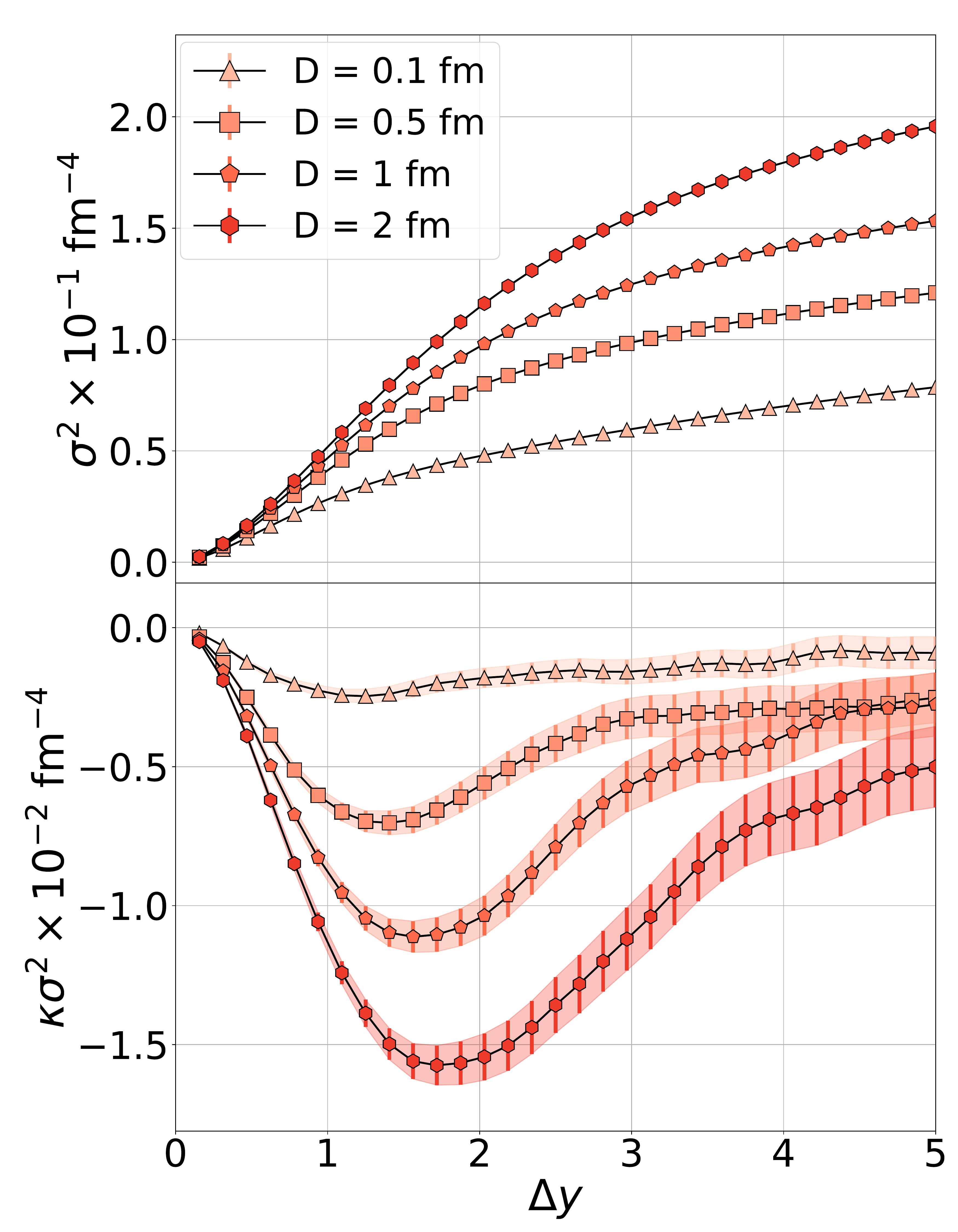}
	\caption{(Color online) The second (upper panel) and fourth (lower panel) order integrated cumulants as a function of the rapidity window $\Delta y$ at $T=145$~MeV and $\mu_B=350$~MeV for different values of the diffusion length $D$. The error bands highlight statistical uncertainties for $2.25 \times10^6$ noise configurations.} 
 \label{fig:DVarKurtD}
\end{figure}

Finally, we show the same rapidity-window dependence of the cumulants for different values of the diffusion length $D$ at $T_f$ and for $\mu_B = 350$~MeV in Fig.~\ref{fig:DVarKurtD}. Again, we observe that both cumulants are strongly impacted by the choice of the diffusion length $D$. For the fourth-order cumulant the non-monotonic structure may be completely lost for small values of $D$, even for the most critical trajectories. In this case, the rapid expansion would wash out the critical signals even qualitatively. We note that for the value of $D=1$~fm applied mostly in this work the critical signal survives the rapid expansion of the medium. It is also interesting to observe that for increasing diffusion length the minimum moves to larger distances in rapidity. So while the signal becomes stronger it would be essential for the experimental set-up to cover a wide range in rapidity to see the non-monotonicity.

\section{Conclusions}
 \label{sec:Conclusions}

In this work we presented a dynamical treatement of the critical fluctuations of the net-baryon density in the context of heavy-ion collisions. We formulated the stochastic diffusion equation for a system undergoing a longitudinal Bjorken expansion. During this expansion the system evolves along different trajectories in $\mu_B$. At $\mu_B=0$~MeV, there is no critical contribution and this trajectory serves as the reference. Increasing $\mu_B$, the critical point is approached and the criticality increases. This behavior in the phase diagram is characterized by a free-energy density functional. Via the second and the fourth-order susceptibilities it contains a critical contribution in the scaling region, which is obtained from the $3$D Ising model, while the regular contributions far from the critical point are constructed in line with lattice QCD calculations.

First, we showed the accurateness of our numerical implementation by comparing the numerical results to analytical discretized expectations in the Gaussian approximation. We demonstrated that the continuum solution is approached as the lattice spacing is decreased. The structure factor and the equal-time correlation function are properly reproduced. We saw the impact of net-baryon charge conservation on the correlation function leading to significant anti-correlations at intermediate rapidities. This benchmarking step is very important to ensure that the inclusion of the stochastic noise term is done correctly.

Then, by using the full nonlinear model we have access not only to the variance but also to the fourth-order cumulant of the fluctuations. We could show that during the time evolution critical signals develop for all critical trajectories and that they are larger the closer the critical point is approached. Compared to the input equilibrium susceptibilities we see that the dynamical evolution broadens the critical signal and moves the results for the different $\mu_B$-trajectories further apart. As experimentally we have access to only one value of this time evolution, corresponding to the freeze-out conditions of the fluctuation observable, we showed the increase of the critical signal as a function of $\mu_B$ for given temperatures. As the temperature is lowered the increase is seen only very close to the most critical trajectory.

In our model the diffusion length is an input parameter, which we have chosen to be of the typical length scale in the QGP, $D=1$~fm. By changing this parameter we can study the interplay between the expansion and the diffusion and especially its impact on the critical fluctuations. We observe that the diffusion length has a strong impact on the cumulants of the net-baryon density fluctuations due to its relation to the relaxation time of the fluctuations. If the diffusion length is significantly shorter than the length scale associated with the expansion, local fluctuations cannot diffuse to form long-range correlations and the anti-correlations, which originate from charge conservation, are equally trapped at short distances in rapidity. This decreases the size of fluctuation observables. For larger diffusion lengths the (anti-)correlations are spread over larger distances and, for example, the correlation function approaches the form in a static system. The diffusive properties of the QGP and the hadron gas are thus of major significance for critical point studies. 

Finally, we have studied the dependence of the second- and the fourth-order cumulants on the rapidity window. While the variance increases for larger rapidity windows and plateaus at half the system size, the fourth-order cumulant shows a pronounced minimum between $\Delta y = 1-2$. The amplitude of this minimum grows as we approach the critical point. The reference trajectory at $\mu_B=0$ does not show this non-monotonicity, which is thus a robust signal of criticality.

In the future we want to couple the diffusion of the net-baryon density fluctuations to fluctuations in the fluid dynamical fields, i.e.~energy density and momentum density. In order to quantify the two main physics consequences of this work - look out for anti-correlations of baryons and the non-monotonic behavior of the fourth-order cumulant as a function of rapidity window - we plan to include a particlization scheme such as in~\cite{Pradeep:2022mkf} and provide calculations for actual particle numbers.

\section*{Acknowledgments}
G.~Pihan, M.~Bluhm and M.~Nahrgang acknowledge the support of the program ``Etoiles montantes en Pays de la Loire 2017''. The authors acknowledge the support of the TYL-FJPPL program of IN2P3-CNRS and KEK. This research was supported in part by the ExtreMe Matter Institute (EMMI) at the GSI Helmholtzzentrum f\"ur Schwerionenforschung, Darmstadt, Germany and by JSPS KAKENHI Grant Numbers 19H05598, 20H01903, 22K03619.

\appendix
\section{Numerical implementation\label{sec:AppA}}

In this Appendix, we provide details on the specific discretization scheme used for our numerical calculations and on the associated analytic expressions in discretized space. In this work, we use a simple Euler forward in time centered in space scheme for the discretization of Eq.~\eqref{eq:FullSDE} which reads 
\begin{multline}
	n_j^{i+1} = n_j^i + \dfrac{D n_c }{ \tau_i \chi_2(\tau_i)} \frac{{\delta \tau}}{{\delta y}^2} \Big(n_{j+1}^i - 2 n_j^i + n_{j-1}^i \Big)\\
        - \frac{D n_c K(\tau_i) }{\tau_i }\frac{{\delta \tau}}{{\delta y}^4}\Big(n_{j+2}^i - 4 n_{j+1}^i + 6 n_j^i - 4 n_{j-1}^i + n_{j-2}^i \Big)\\ 
	+ \dfrac{D \lambda_4(\tau_i)}{n_c \tau_i}\frac{{\delta \tau}}{{\delta y}^2}\Big(\left(n_{j+1}^i\right)^3 - 2\left(n_j^i\right)^3 + \left(n_{j-1}^{i}\right)^3 \Big)\\
	- \sqrt{\dfrac{2 D n_c {\delta \tau}}{\tau_i  {\delta y}^3 }}\Big(W_{j+1}^i - W_{j}^i \Big) \,.
	\label{eq:DisSDE}
\end{multline}
Here, the lower index $j$ accounts for the site of cell $j$ in space-time rapidity $y$ and the upper index $i$ for the time-step $i$ in the temporal evolution, such that $\tau_i=\tau_0+i\cdot\delta\tau$. Accordingly, $n_j^i$ quantifies the cell-average of the net-baryon density fluctuations $\widetilde n_B(\tau_i,y_j)$ at $\tau_i$ and cell-position $y_j=-L/2+j\cdot\delta y$. For the increment in proper-time $\delta\tau$ we choose $\delta\tau=0.2\,(\delta y)^4 n_c/(8KD)$ for stability reasons, while the resolution in space-time rapidity is defined as $\delta y=L/N$ for $N$ sites. Throughout this work we consider $L=20$ units of space-time rapidity for the total size of the system. Finally, $W_j^i$ represents a centered Gaussian white noise with unit variance at $\tau_i$ and $y_j$. 

In Eq.~\eqref{eq:DisSDE}, all terms have an explicit proper-time dependence. This makes, even in the Gaussian approximation with $\lambda_4=0$, an analytic calculation of the structure factor at equal proper-time $\tau_i$ very difficult. For this reason, we instead choose to consider Eq.~\eqref{eq:FullSDE} for discretized space-time rapidity but continuous proper-time which reads 
\begin{multline}
	\partial_{\tau} n_j(\tau_i)=  \dfrac{D n_c }{ \tau_i \chi_2(\tau_i)} \frac{1}{{\delta y}^2}\Big(n_{j+1}(\tau_i) - 2 n_j(\tau_i) + n_{j-1}(\tau_i) \Big)\\
        -\frac{D n_c K(\tau_i) }{\tau_i }\frac{1}{{\delta y}^4}\Big(n_{j+2}(\tau_i) - 4 n_{j+1}(\tau_i) + 6 n_j(\tau_i)\\ 
	- 4 n_{j-1}(\tau_i) + n_{j-2}(\tau_i) \Big)\\ 
	+ \dfrac{D \lambda_4(\tau_i)}{n_c \tau_i}\frac{{1}}{{\delta y}^2}\Big({n_{j+1}(\tau_i)}^3 - 2 {n_j(\tau_i)}^3 + {n_{j-1}(\tau_i)}^3 \Big)\\
	- \sqrt{\dfrac{2 D n_c}{\tau_i {\delta y}^3}}\Big(W_{j+1}(\tau_i) - W_{j}(\tau_i)\Big) \,.
\label{eq:DisSDEtau}
\end{multline}
From Eq.~\eqref{eq:DisSDEtau} in the Gaussian approximation with $\lambda_4=0$ a formal solution for $n_q(\tau_i)$ equivalent to Eq.~\eqref{eq:FormSol} but for discrete values of $q$ can be obtained after spatial Fourier transformation. From this the structure factor at equal proper-time $S_q(\tau_i)$ is calculated and found to have the same form as $S(\tau_i,q)$ in Eq.~\eqref{eq:Sk} with coefficients $\alpha(q)$, $\beta(q)$ and $\gamma(q)$ given in Eq.~\eqref{eq:CoeffDis}. It is $S_q(\tau_i)$ which is presented in Fig.~\ref{fig:Sk150} as analytic solutions in discretized space by dashed lines. The almost perfect agreement with our numerical results shows that with the chosen proper-time increment $\delta\tau$ we are already sufficiently close to the temporal continuum limit. We have numerically verified that this is indeed the case by varying $\delta\tau$ within the stability range.

In the numerics, one may exploit Eq.~\eqref{eq:DisSDE} in order to prepare equilibrated initial conditions prior to the rapid cooling and expansion as used in section~\ref{sec:Dynamics}. This is accomplished by iterating Eq.~\eqref{eq:DisSDE} for sufficiently many numerical steps without actually advancing the proper-time argument from $\tau_0$ and lowering the temperature from $T_i$. 

Only after equilibrium is reached, which may be verified from the iteration-step independence of local fluctuation observables or the correlation function for example, the dynamical evolution in line with Eq.~\eqref{eq:Hubble} is switched on.

\bibliographystyle{ieeetr}
\bibliography{biblio}

\end{document}